\documentclass[aip,reprint,footinbib,hidelinks]{revtex4-1}
\usepackage{tabularx}
\usepackage{graphicx}
\usepackage{dcolumn}
\usepackage{color}
\usepackage{soul}
\usepackage{bm}
\usepackage{hyperref}
\hypersetup{
    colorlinks=false,
    linkcolor=black,
    filecolor=black,      
    urlcolor=black,
}

\begin{document}

\title{Integration of multi-layer black phosphorus into photoconductive antennas for THz emission} 
\author{M.~H.~Doha}
\affiliation{Department of Physics, University of Arkansas, Fayetteville, AR 72701}
\author{J.~I.~Santos Batista}
\affiliation{Department of Electrical Engineering, University of Arkansas, Fayetteville, AR 72701}
\author{A.~F.~Rawwagah} 
\affiliation{Department of Physics, University of Arkansas, Fayetteville, AR 72701}
\author{J. P.~Thompson}
\affiliation{Department of Physics, University of Arkansas, Fayetteville, AR 72701}
\author{A. Fereidouni}
\affiliation{Department of Physics, University of Arkansas, Fayetteville, AR 72701}
\author{K.~Watanabe}
\affiliation{Research Center for Functional Materials, National Institute for Materials Science, 1-1 Namiki, Tsukuba 305-0044, Japan}
\author{T.~Taniguchi} 
\affiliation{International Center for Materials Nanoarchitectonics, National Institute for Materials Science, 1-1 Namiki, Tsukuba 305-0044, Japan}
\author{M.~El-Shenawee}
\affiliation{Department of Electrical Engineering, University of Arkansas, Fayetteville, AR 72701}
\author{H.~O.~H.~Churchill}
\email{churchill@uark.edu}
\affiliation{Department of Physics, University of Arkansas, Fayetteville, AR 72701}

\date{\today}

\begin{abstract}
We report the fabrication, characterization, and modeling of photoconductive antennas using 40 nm thin-film flakes of black phosphorus (BP) as the photoconductor and hexagonal boron nitride (hBN) as a capping layer to prevent oxidation of BP. Dipole antennas were fabricated on oxidized high-resistivity Si substrates, and BP and hBN flakes were picked up and transferred onto the antenna inside a nitrogen glovebox. The transfer matrix technique was used to optimize the thickness of BP and hBN for maximum absorption. BP flakes were aligned with the armchair axis along the anode-cathode gap of the antenna, with crystal orientation measured using reflection anisotropy. Photocurrent imaging under illumination with 100 fs pulses at 780 and 1560 nm showed a bias-dependent maximum photocurrent localized to the antenna gap with a peak photoconductivity of 1 (2) S/cm in the linear regime of bias for excitation at 780 (1560) nm. Photocurrent saturation in bias (pump fluence) occurred at approximately 1 V (0.25 mJ/cm$^2$).  Device performance was modeled numerically by solving Maxwell's equations and the drift-diffusion equation to obtain the photocurrent density in response to pulsed laser excitation, which was largely in qualitative agreement with the experimental observations. THz output computed from surface current density suggests that BP THz PCA performance is at least comparable to more traditional devices based on low-temperature-grown GaAs. These devices represent a step toward high-performance THz photoconductive antennas using BP.\end{abstract}


\maketitle 

\section{Introduction}
Terahertz photoconductive antennas (PCAs) generate broadband THz pulses when a photoconductor is excited by short ($\sim$100 fs) laser pulses focused to the gap between the anode and cathode of the antenna.
These THz pulses, with bandwidths typically covering the range of 0.1 to 4 THz, are commonly used in commercial systems for THz time-domain spectroscopy.
However, PCAs based on the standard material, low-temperature-grown GaAs, suffer from low power conversion efficiency, typically in the range of 10$^{-5}$ to 10$^{-4}$ (Ref.~\onlinecite{burford2017review}).
Several approaches have been implemented recently to improve the performance of GaAs THz PCAs, including the use of plasmonic elements to enhance absorption,\cite{berry2013significant,burford2017plasmonic} thin GaAs to prevent THz absorption loss within the photoconductor,\cite{burford2017plasmonic} and nanostructuring to reduce photocarrier transit time.\cite{yardimci2017high,yardimci2018high}
Incorporation of three-dimensional plasmonic elements at the PCA electrodes has produced an optical-to-THz conversion efficiency of 7.5\%.\cite{yang20147}

The anisotropic, layered semiconductor black phosphorus (BP)\cite{xia2014rediscovering,liu2014phosphorene,koenig2014electric,qiao2014high} combines several attractive properties for application as the photoconductor in THz PCAs:  strong absorption at the wavelengths of low-cost fs fiber lasers (780 and 1560 nm),\cite{lan2016visualizing} high room-temperature hole (electron) mobility of 5,000 (600) cm$^2$/Vs,\cite{long2016achieving,akahama1983electrical} and few-layer BP doped with Al adatoms was reported to have an electron mobility of 1,500 cm$^2$/Vs near room temperature.\cite{prakash2017black}
Importantly for THz PCA applications, BP also supports a high saturation velocity of 1.5 $\times$ 10$^7$ cm/s.\cite{li2019high}

Because the excitation pulse in a PCA is localized to the vicinity of a few-micron gap in the antenna, the small size of exfoliated BP flakes is not a limitation for this application.
Furthermore, the small lateral size of flakes of mechanically exfoliated BP is advantageous to prevent THz absorption loss that occurs when using a bulk photoconductor, and van der Waals-based dry transfer permits integration with low-loss antenna substrates such as high-resistivity Si.

The optoelectric properties of BP have been widely investigated, with photoresponse dominated by photovoltaic (bolometric) photocurrent at low (high) doping.\cite{low2014origin,hong2014polarized,buscema2014photovoltaic,buscema2014fast,deng2014black,miao2019black}
For THz applications, BP-based THz detectors have been characterized,\cite{viti2015black,viti2019thermoelectric} including integration of BP into a PCA for THz detection.\cite{mittendorff2017optical}
However, the superior material properties of BP for THz PCAs described above suggest that BP could also form the basis of efficient, broadband THz sources.
Here we describe a significant step in this direction by integrating multi-layer BP as the photoconductor in PCAs and characterizing the photoresponse of the devices in the high-bias, high-fluence regime relevant to THz PCA emitters.

Two-dimensional material flake thicknesses and substrates chosen to maximize absorption for optoelectronic applications have previously been reported in Refs. \onlinecite{song2015nanocavity,bahauddin2016broadband,wang2017fabry,li2019engineering,jariwala2016near}, in addition to more complex architectures that, for example, take advantage of the coupling between a cavity and relatively narrowband excitonic transitions.\cite{doi:10.1021/acs.nanolett.0c00492}
Using an optical model for the reflection, absorption, and transmission of light within the device, we selected thicknesses of BP and encapsulating hexagonal boron nitride (hBN) layers to form an anti-reflection cavity that optimizes absorption in the BP layer. We then characterized the photoresponse of the BP PCAs using scanning photocurrent microscopy.
We further modeled the devices computationally by solving Maxwell's equations and the drift-diffusion equation to obtain the optical absorption and photocurrent density in response to pulsed laser excitation.
The experimentally observed strong absorption of 36\% at 780 nm, photocurrent saturation in bias (pump fluence) above 1 V (0.25 mJ/cm$^2$), and peak photoconductivity of 1 (2) S/cm at 780 (1560 nm) suggest a promising outlook for the THz performance of these devices, though further work may be needed to engineer a short carrier lifetime in BP\cite{wang2016ultrafast} or minimize device transit time.

\section{Experimental and Computational Methods}
\subsection{Experimental Methods}
A schematic of the BP PCA devices we fabricated is shown in Fig.1(a), in which BP was covered by an hBN layer and contacted on the bottom with Au. Fabrication began with electron beam lithography to define the dipole antenna on a Si substrate with resistivity $>10,000$ $\Omega$-cm and 300 nm of thermal oxide [Fig.~1(c)].
After developing the ebeam resist and prior to metal deposition, the chip was etched for 36 seconds in 5:1 buffered oxide etch to planarize the contact surface with respect to the oxide (cf.~supplemental Fig.~S1) and prevent damage to the BP during transfer onto the antenna.
The etch was followed by thermal evaporation of 5/60 nm of Cr/Au. 
BP and hBN were mechanically exfoliated from bulk crystals, and flakes with thicknesses chosen to maximize absorption at 780 nm (see below) were identified using a hermetic cell with optical access.\cite{thompson2019exfoliation} Flake thicknesses were confirmed using atomic force microscopy inside a nitrogen glovebox (H$_2$O, O$_2$ $<$ 0.5 ppm).
Flakes were transferred onto the antenna using polycarbonate-based dry transfer, also inside the glovebox.\cite{zomer2014fast}
\begin{figure}
\includegraphics[width=3.37in]{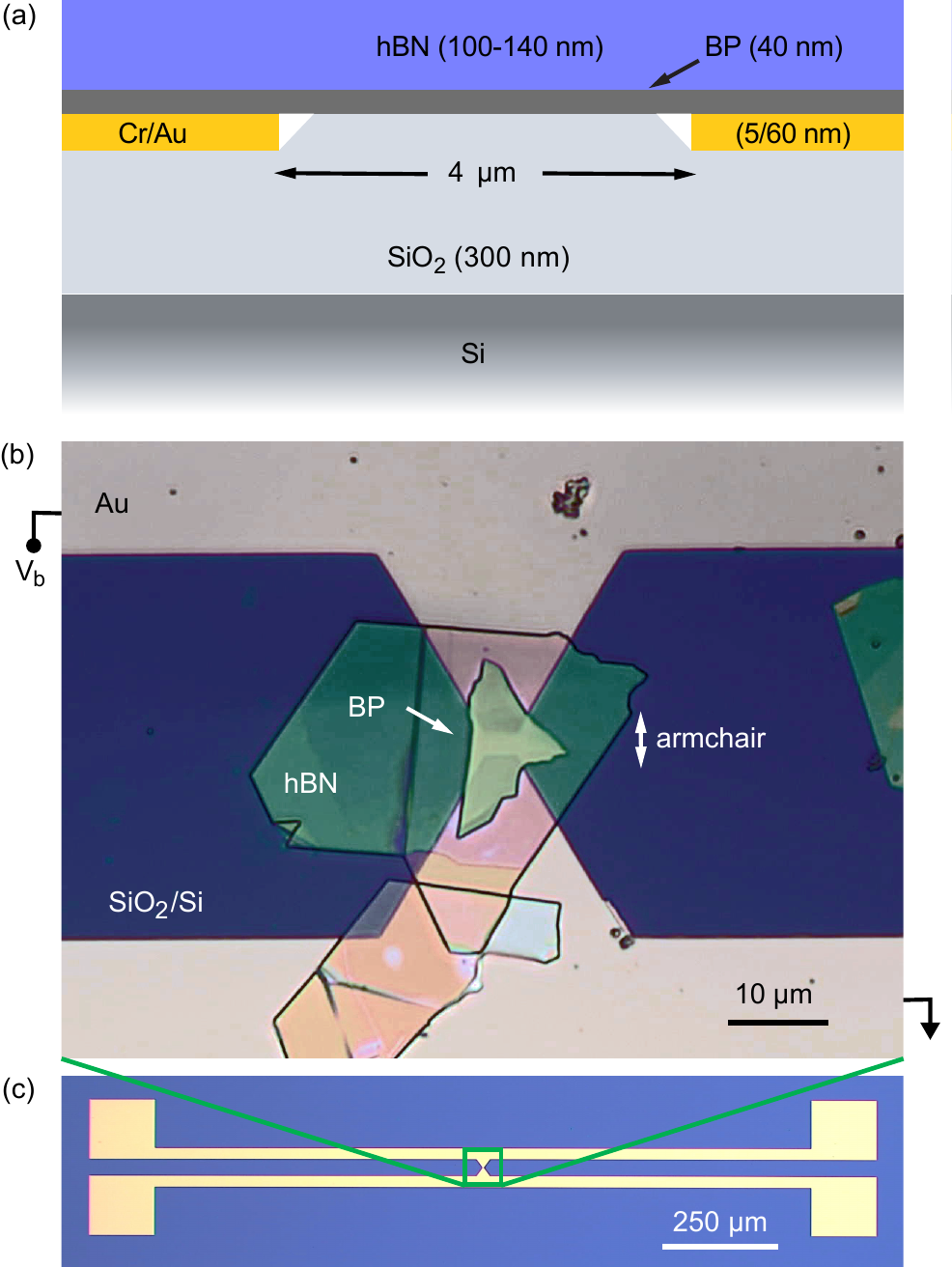}%
\caption{\label{fig1}(a) Cross-section of BP photoconductive antenna in which BP encapsulated by hBN is transferred onto recessed Au contacts that are planarized with the Si/SiO$_2$ substrate. (b) Optical micrograph of a complete device in which encapsulated BP bridges the gap of a dipole antenna. (c) Optical micrograph of the antenna.}%
\end{figure}

Devices were characterized at room temperature by measuring AC photocurrent in response to excitation by 100 fs pulses with a repetition rate of 80 MHz at 1560 nm from an Er-doped fiber laser, with a DC bias $V_b$ applied to one side of the antenna [Fig.~1(b)].  
For measurements at 780 nm, the pulses were frequency doubled using a periodically-poled LiNbO$_3$ crystal and passed through a bandpass filter centered at 780 nm with a bandwidth of 10 nm.
AC measurements used a mechanical chopper at 811 Hz, a current preamplifier, and a lock-in amplifier.
The polarization of the excitation beam was rotated using a half-wave plate to align with the armchair axis of the BP flakes to maximize photocurrent.
Photocurrent and reflectance images were acquired simultaneously using a two-axis galvo-mirror scanner driven by a homemade synchronized digital-to-analog and analog-to-digital converter pair\footnote{\url{http://www.opendacs.com}} operating at a line-scan frequency of approximately 5 Hz with each line averaged 10 times.
During measurements, the BP PCAs were maintained in a nitrogen environment using a hermetic cell similar to that described in Ref.~\onlinecite{thompson2019exfoliation}.
Under these conditions, no environmental degradation of the BP devices was observed during measurements over at least eight weeks.

To maximize the efficiency of BP PCAs, we leverage the anisotropic mobility, saturation velocity, and optical absorption of BP.
Differential reflectance anisotropy using 780 nm excitation was used to determine BP crystal axis orientation prior to transfer so that the high-velocity, strong absorption armchair axis could be aligned with the electric field created by the bias applied to the antenna. 

\subsection{Computational Methods}
An efficient PCA must be strongly absorbing at the wavelength of the pump. 
We used the transfer matrix method to calculate the absorption by the hBN/BP/SiO$_2$/Si stack as a function of hBN and BP thickness for light at 780 nm normally incident on the device.\footnote{Source code available at\newline \url{https://github.com/churchill-lab-ar/RAT}.} 
We used this model to calculate absorption in different regions of the PCA, with excitation for example inside the antenna gap or on the antenna anode or cathode and under the valid assumption that the excitation wavelength is much smaller than the size of the antenna gap.
The reported values in Refs.~\onlinecite{schuster2015anisotropic} and \onlinecite{lee2019refractive} were used for the complex refractive indices for BP and hBN, respectively.
Flake thicknesses were restricted to be greater than 10 nm to permit the use of bulk optical constants over the entire calculation range. 
Because of the strongly anisotropic optical constants of BP, we performed the calculation for light polarized along both the armchair and zigzag axes of BP.

The tangential components of electric and magnetic fields $\textbf{E}_a$ and $\textbf{H}_a$ at the interface between vacuum and the hBN/BP/SiO$_2$ stack are continuously connected with $\textbf{E}_b$ and $\textbf{H}_b$ transmitted through the final interface between SiO$_2$ and Si via\cite {macleod2017thin}
\begin{equation}
\left[ \begin{array}{c} B \\ C \end{array} \right] = \left\{ \prod_{r = 1}^{q} \left[ \begin{array}{cc} \cos\delta_r & (i \sin\delta_r)/\eta_r \\ i \eta_r \sin\delta_r & \cos\delta_r \end{array} \right]\right\} \left[ \begin{array}{c} 1 \\ \eta_m \end{array} \right], \label{eq1}
\end{equation}
where B = $\textbf{E}_a$/$\textbf{E}_b$ and C = $\textbf{H}_a$/$\textbf{E}_b$ are the normalized electric and magnetic fields at the front interface, $\delta_r = 2\pi \tilde{n}_r d \cos\theta/\lambda$ is the phase thickness of the film, $\tilde{n}_r = n_r - i \kappa_r$ is the complex refractive index, $n_r$ is the refractive index, $\kappa_r$ is the extinction coefficient, d is the film thickness, $\theta$ is the incident angle, $\lambda$ is the wavelength, $y = \sqrt{\epsilon_0/\mu_0}$ is the free space optical admittance, $\eta^s_r = y \tilde{n}_r \cos\theta_r$ is the admittance for s-polarized light, $\eta^p_r = y \tilde{n}_r/\cos\theta_r$ is the admittance for p-polarized light, $\eta_0$ is the vacuum admittance, and $\eta_m$ is the substrate admittance. The absorption was calculated using the surface admittance $Y$, which is the ratio of the total tangential magnetic and electric fields at the incident interface and is given by $Y = \textbf{H}_a/\textbf{E}_a = C/B$.
Finally, the fraction of light absorbed within the stack was calculated as
\begin{equation}
    A = \frac{4 \eta_0 Re(B C^* - \eta_m)}{(\eta_0 B + C) (\eta_0 B + C)^*}.
\end{equation}

\begin{figure}
\includegraphics[width=3.37in]{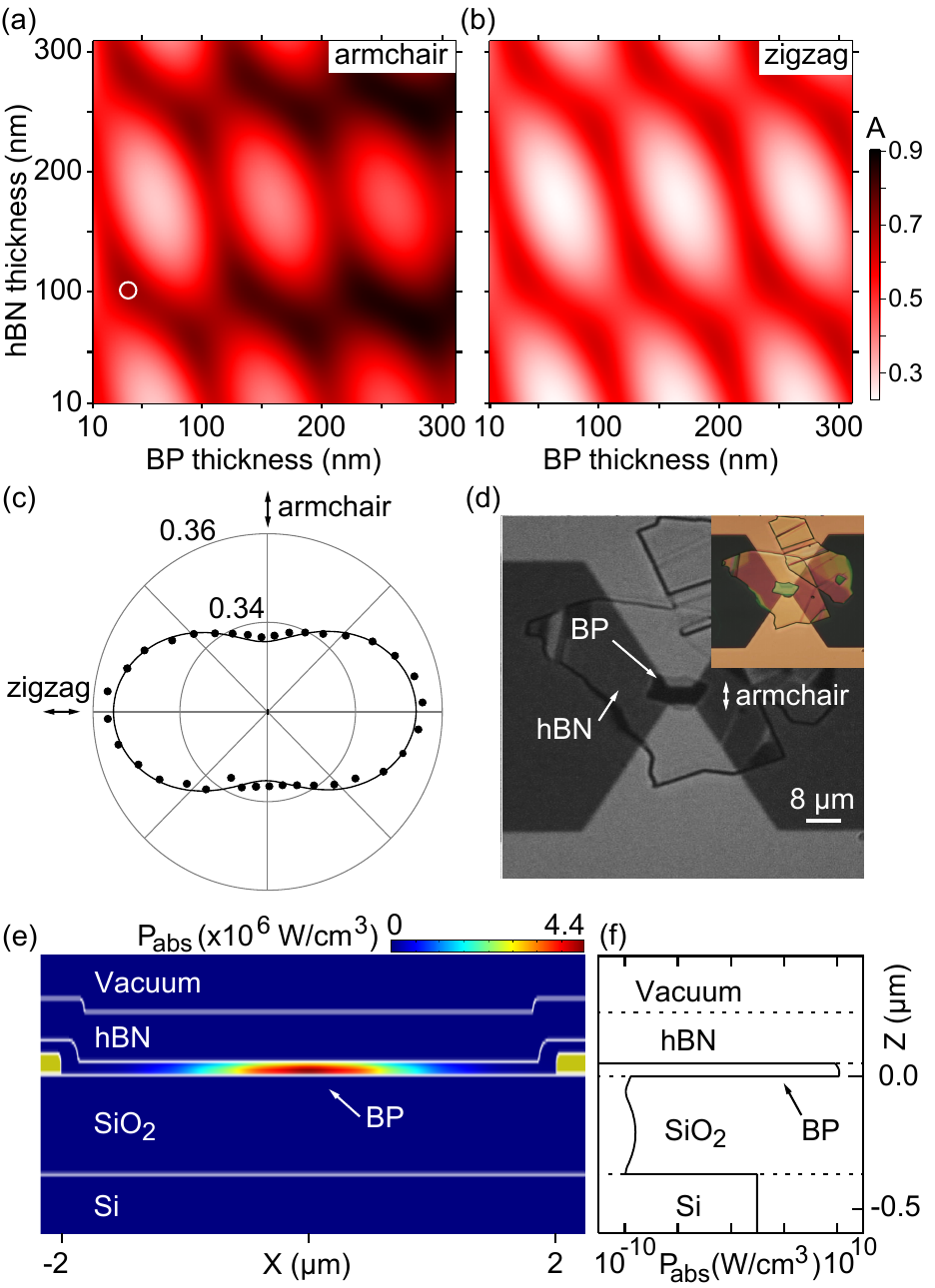}
\caption{\label{fig2}Absorption of light at 780 nm polarized along the (a) armchair and (b) zigzag axis of BP by the hBN/BP/SiO$_2$/Si stack as a function of hBN and BP thickness calculated using the transfer matrix method.  The white circle indicates the selected flake thicknesses for our devices with BP thickness of 40 nm and hBN thickness of 100 nm. (c) Circles: reflectance of a BP flake on a SiO$_2$/Si substrate as a function of polarization of the 780 nm probe beam. Solid line: sinusoidal fit to the reflectance.  (d) Reflection image (unpolarized, red-channel only) of a BP PCA, filtered at 780 nm.  Inset: White-light reflection image of the same device. (e) Calculated average absorbed power, $P_{\rm abs}$, within each layer of the device.  (f) Line cut of (e) at $X = 0.$}
\end{figure}

To further understand absorption and photocurrent generation in BP PCAs, we also modeled devices using a computational COMSOL multiphysics model that solved Maxwell's equations to obtain the electric and magnetic fields in response to the pulsed laser excitation. The computational domain of the PCA consisted of layers of vacuum, hBN, BP, SiO$_2$, and Si as shown in Fig.~1(a). This solution is based on the use three modules of the computational COMSOL multiphysics package (optical, semiconductor, and antenna modules). In this model we used the absorbing scattering boundary condition around the computational domain. The optical module solves the wave equation in the frequency domain. The electric and magnetic fields $\textbf{E}(r,t)$ and $\textbf{H}(r,t)$ are obtained in response to the laser excitation at 780 (1560) nm. Based on the laser average incident power, the incident electric field is given by Ref.~\onlinecite{burford2016computational}:
\begin{equation}
    \mathbf{E}_{inc} = \hat{x}\sqrt{\frac{4\eta_{0} P_{ave}}{\pi D_x^{2}}}\exp\left[4\ln(0.5)\frac{(x-x_{0})^{2}}{D_{x}^{2}}\right]
\end{equation}
where $P_{ave}$ is the power average of a train of laser pulses, $x_{0}$ is the center of the laser position, and $D_{x}$ is the Gaussian half-power beam width along $\hat{x}$. In this module, the BP was treated as an anisotropic material with complex permittivity and conductivity in the x-, y- and z-directions. The laser excitation is polarized in the x-direction which is the armchair direction shown in Fig.~2.

The electric and magnetic fields in the BP layer were extracted from the optical module to compute the maximum power density $P_{max}(r)$ at each position point $(r)$ in the layer. We observed that the power in the z-direction was several orders of magnitude higher than that in the y- and x-directions, which is expected based on the incident polarization in equation (3). In the semiconductor module, the power density $P_{max}(r)$ is multiplied by the Gaussian function to account for the amplitude modulation of the laser excitation to obtain the carrier generation function $R_{g}(r,t)$ as Ref.~\onlinecite{burford2016computational}:
\begin{equation}
    R_{g}(r,t) = \frac{4\pi\kappa P_{max}(r)}{hc}\exp\left[4\ln(0.5)\frac{(t-t_{0})^{2}}{D_{t}^{2}}\right]
\end{equation}
where $\kappa$ represents the extinction coefficient of the photoconductive material, $h$ is Planck's constant, $c$ is the speed of light, $t_0$ is the time center of the laser pulse, and $D_t$ is the laser pulse width in time. For the carrier recombination rate, we used the Shockley–Read–Hall model.\cite{lien2009physics}

The semiconductor module solves the standard drift-diffusion and Poisson's equations in the time domain. However, this module does not account for the anisotropy of the BP material. As a result we treated the material as isotropic with the complex permittivity of the x-direction (armchair direction) which is also the direction of the bias voltage of the device and the laser polarization. The solution of the photocurrent density $\textbf{J}(r,t)$ is the summation of the electron and hole currents $\textbf{J}_n(r,t)$ and $\textbf{J}_p(r,t)$, respectively, given by \cite{selberherr1984physical}:
\begin{equation}
    \begin{array} {lcl}
        \mathbf{J}_{n,p}(r,t) &=& \mu_{n,p} \nabla E_{c,v} m(r,t) \\
        & & \pm \mu_{n,p} k_{B} TF\left(\frac{m(r,t)}{N_{c,v}}\right)\nabla m(r,t)
    \end{array}
\end{equation}
where $m(r,t)$ is $n(r,t)$ or $p(r,t)$ for $\mathbf{J}_n(r,t)$ or $\mathbf{J}_{p(r,t)}$, respectively, $\mu_{n}$ and $\mu_{p}$ are the electron and hole mobility, $E_{c}$ and $E_{v}$ are the conduction and valence band energies, $N_{c}$ and $N_v$ are the effective density of states for electrons and holes, $F$ is the Fermi function, $T$ is the temperature, $k_{B}$ is the Boltzmann constant, and the positive (negative) sign in the second term applies to electron (hole) photocurrent. Here the temperature $T$ was assumed constant as room temperature, therefore our results did not include thermal effects in the device. To account for  photocurrent saturation, the carrier mobility is assumed to vary with the induced electric field based on the Caughey-Thomas mobility model.\cite{caughey1967carrier}

In this model, the carrier mobility of holes (electrons) in BP is assumed to be 5,000 (1,500) cm$^{2}$/Vs, to represent an upper limit of BP PCA performance.\cite{long2016achieving,prakash2017black} 
The carrier lifetime of BP has been reported to be frequency dependent for excitation in the near infrared, and we take lifetimes of 0.36 ps and 0.93 ps at 780 and 1560 nm, respectively.\cite{wang2016ultrafast} 
To match parameters of the device for which photoresponse data are presented, we also took an hBN thickness of 140 nm, BP thickness of 40 nm, SiO$_{2}$ thickness of 300 nm, and assumed semi-infinite Si. 
Complex refractive indices for BP and hBN were taken to be the same as those used in the transfer matrix model with the BP complex relative permittivity being $16.06 - 1.73i$ ($14.48 - 2.23i$) in the x-direction and $14.76 - 0.096i$ ($13.41 - 0.039i$) in the y-direction for 780 (1560) nm, respectively.
The BP permittivity in the z-direction is 8.3 for both frequencies. \cite{nagahama1985optical}  We take the conductivity of BP to be 2.5 S/cm in the x-direction, 0.93 S/cm in the y-direction, and 0.44 S/cm in the z-direction. \cite{akahama1983electrical} The Si and SiO$_{2}$ have complex permittivities of $13.51 - 0.04i$ ($11.91 - 0.13i$) and 2.38 (2.36), respectively, at wavelengths of 780 (1560) nm.\cite{green2008self,ghosh1999dispersion,pierce1972electronic}

The photocurrent density $\mathbf{J}(r,t)$ is extracted from the semiconductor module to provide a current source in the antenna module. In the antenna module, the hBN, BP, and SiO$_{2}$ layers were ignored due to their insignificant thickness compared with the THz wavelength. The generated THz electric field signal $\mathbf{E}_{THz}(r,t)$ can be obtained at any point in the domain as \cite{jackson1975classical}:
\begin{equation}
    \textbf{E}_{THz}(r,t) = -\frac{\mu_0}{4\pi}\frac{\partial}{\partial t}\int{\frac{\mathbf{J}_{s}(r,t-(|r-r'|/c))}{|r-r'|}ds'}
\end{equation}
where $\mu_{0}$ is the magnetic permeability, $\mathbf{J}_{s}(r,t)$ is the photoconductive surface current at the antenna gap, $|r-r'|$ is the distance between the source and the field point, and $ds'$ is the increment of surface area at a displacement $r'$ from the antenna center. Collectively based on the above three modules, the generated THz electric field signal solution is function of the optical response, carrier mobility, carrier saturation velocity, carrier recombination, bias voltage, laser power and wavelength, carrier lifetime, and photoconductor permittivity and conductivity. More details of the computational model were reported for GaAs PCA in Ref.~\onlinecite{burford2017plasmonic}.

\section{Results and Discussion}

The results of the transfer matrix calculation [Fig.~2(a),(b)] indicate strong absorption by the device when the BP and hBN thicknesses create an antireflective quarter-wave cavity, with a weaker dependence on the hBN thickness because of its much smaller dielectric constant compared to BP.
Among many strongly absorbing BP thicknesses to choose from, we prefer those with thinner BP to retain the possibility of applying a gate voltage to adjust the carrier density of BP in future devices, which will allow us to minimize dark current and maximize breakdown electric field.
As indicated by the white circle in Fig.~2(a), absorption is maximized with BP approximately 40 nm thick and hBN approximately 100 nm thick.
\begin{figure}
\includegraphics[width=3.37in]{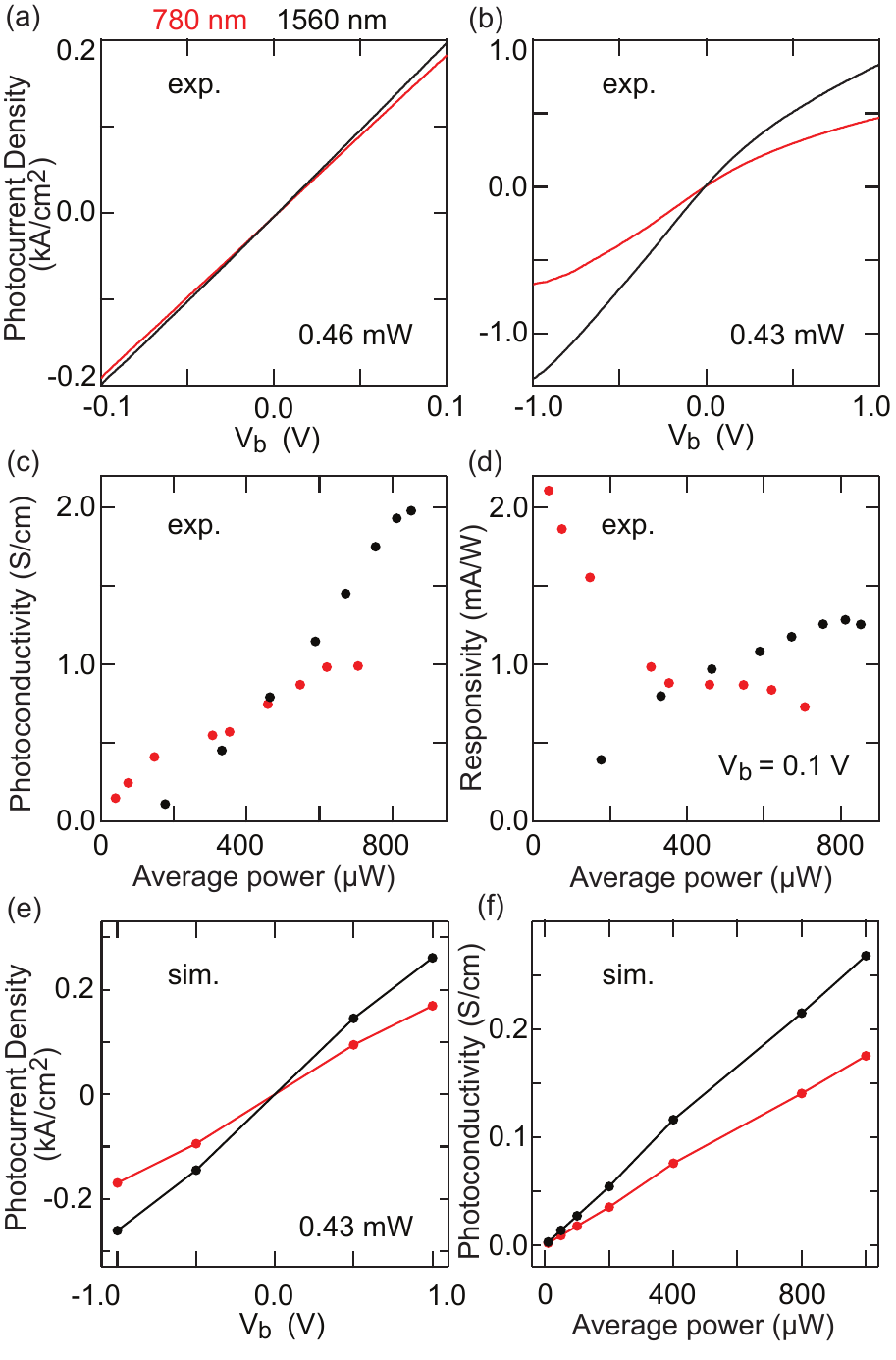}%
\caption{\label{fig3}(a) AC photocurrent as a function of $V_b$ for excitation of the BP PCA at 780 and 1560 nm at an average power of 0.46 mW. In all panels red (black) curves or circles are associated with excitation at 780 (1560) nm.  (b) Higher bias photocurrent at an average power of 0.43 mW.  (c) Photoconductivity measured at $V_b=100$ mV. (d) Responsivity extracted from the data in (c). (e) Simulated photocurrent density as a function of V$_b$ for excitation of the BP PCA at 780 nm (red) and 1560 nm (black) at an average power of 0.43 mW. (f) Simulated photoconductivity as function of average power extracted from photocurrent over the range V$_b$ = $\pm$ 100 mV. }%
\end{figure}
Comparing Fig.~2(a) and 2(b), absorption by the device for light polarized along the armchair direction of BP is approximately 20\% larger than for light polarized along the zigzag direction, justifying our alignment of the excitation beam polarization with the armchair direction of the BP. This observation also allows the use of narrow-band differential reflectance anisotropy measurements to determine the crystal orientation of BP flakes.
A differential reflectance anisotropy measurement of a device with a 40 (100) nm-thick BP (hBN) flake is shown in Fig.~2(c).
Light is reflected more weakly for polarization aligned to the strongly absorbing armchair axis of the BP, and we use this information to transfer the BP flake onto the antenna with the armchair axis aligned to the electric field produced by $V_b$ to within 2-3$^{\circ}$.

A reflection image bandpass-filtered around 780 nm is shown in Fig.~2(d), in which we observe a dark region covering the antenna gap that corresponds to the strongly absorbing BP layer.
Additionally, the hBN flake in Fig.~2(d) is transparent to light at 780 nm, indicating that absorption by BP is maximized when the hBN is transparent, as expected.
Using this image we compare the light intensity reflected by the hBN/BP/SiO$_2$/Si region, $R_{BP}$, with the light intensity reflected by the hBN/SiO$_2$/Si region, $R_{hBN}$, to quantify the absorption by BP as $1 - R_{BP}/R_{hBN} = 0.71\pm0.04$, with uncertainty derived from the standard deviation of pixel intensity over the measurement region.
Calculated absorption using the transfer matrix method yields an expected range of absorption 0.69$^{+0.00}_{-0.06}$, assuming an uncertainty in BP and hBN thicknesses of $\pm5$ nm.

To clarify the role of BP in the photoresponse of the fabricated device, we calculated the average absorbed power in each layer using the COMSOL solution of Maxwell's equations for an excitation wavelength of 780 nm. The map shown in Fig.~2(e) indicates that the BP layer dominates absorbed power relative to other layers.
Plotting the average absorbed power along the z-direction in the middle of the antenna gap [Fig.~2(f)], we see that absorption by the BP layer is many orders of magnitude larger than absorption by the Si substrate and that, as expected, absorption by vacuum, hBN, and SiO$_2$ are negligible. 
This result indicates that the primary loss of incident power is from reflection rather than absorption by the Si substrate.
\begin{figure*}
\includegraphics[width=7in]{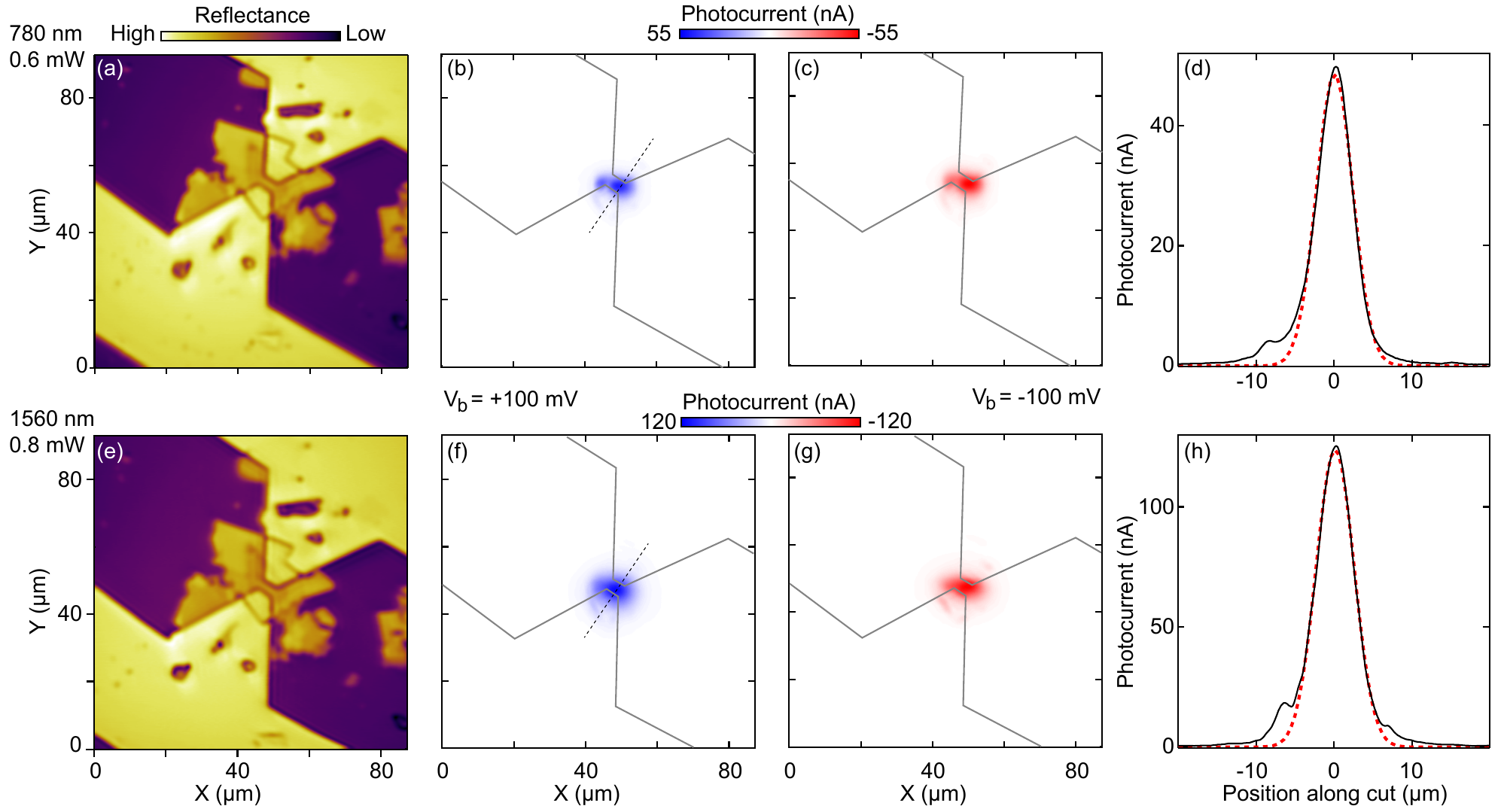}\caption{\label{fig4}Scanning reflectance and AC photocurrent images of a BP PCA.  (a)-(c) Images for excitation at 780 nm. (d) Line profile of photocurrent along the path indicated by the black dashed line in (b), for $V_b=100$ mV.  The red dotted line is a Gaussian fit to the photocurrent profile. (e)-(h) Images for excitation at 1560 nm.  Photocurrent images in (b),(f) were acquired with $V_{b} = 100$ mV and images in (c), (g) were acquired with $V_{b}=$-100 mV. (h) Similar to (d) along the path shown in (f). Gray lines in (b), (c), (f), and (g) indicate the outlines of the contacts based on the reflection images in (a) and (e).}%
\end{figure*}

We characterized the photoresponse of three nominally identical BP PCAs and present the results from one of them.
Data for additional BP PCAs are presented in supplementary Figs.~S2 and S7.
This BP PCA had BP (hBN) thicknesses of 40 (140) nm (Fig.~S3) and a DC dark resistance of 15 k$\Omega$ (Fig.~S5), comparable to devices based on other materials capable of operating at 1560 nm such as InGaAs.\cite{glinskiy2017total}
We estimate that this thicker than optimum hBN layer reduces absorption by approximately 20\% relative to an ideal device [cf.~Fig.~2(a)].
AC photocurrent characteristics of the BP PCA are shown in Fig.~3(a) for illumination by pulses with an average power of 460 $\mu$W at both 780 and 1560 nm.
For both wavelengths the photocurrent is linear at low bias.
At higher bias [Fig.~3(b)] the photocurrent becomes sublinear above approximately $V_b = 0.1$ V and is approximately 50\% larger for negative bias, likely caused by asymmetrical contact resistances between Au and BP at the anode and cathode.
Notably, the high-bias photocurrent is 50\% larger for excitation at 1560 nm than at 780 nm.

Photoconductivity of the device as a function of average power was measured by fitting the slopes of linear photocurrent vs.~bias curves over the range $\pm100$ mV at different powers.
Photoconductivity of the device increases with average power [Fig.~3(c)], saturating at 1 S/cm for excitation at 780 nm.
For 1560 nm excitation, photoconductivity reached 2 S/cm at the highest power available (850 $\mu$W) and appears to begin to saturate at that power.
The irradiance (fluence) at this power is approximately 20 kW/cm$^2$ (0.25 mJ/cm$^2$) for our measured spot size of 2 $\mu$m.

To provide a comparison with the large experimental literature concerning BP photodetectors,\cite{guo2016black,engel2014black,huang2016broadband,wu2015colossal,buscema2014fast,hou2018multilayer,venuthurumilli2018plasmonic,yuan2015polarization,chen2017three,gong2018black,youngblood2015waveguide} we also compute the responsivity of this device as a function of average power at $V_b=100$ mV, as shown in Fig.~3(d).
Strikingly different behavior was observed for 780 and 1560 nm excitation:  whereas responsivity decreases with power for 780 nm excitation, it increases with power for 1560 nm excitation and saturates at a modest value of about 1.25 mA/W.
This observation implies sub(super)-linear photoconductivity as a function of power for excitation at 780 (1560) nm.
This result may indicate superior efficiency of future BP THz PCAs pumped at 1560 nm compared with 780 nm. 
Although very large responsivities have been reported for BP photodetectors,\cite{liu2018highly} such measurements are obtained for average powers/fluences three to four orders of magnitude smaller than reported here.
In contrast, because of the quadratic dependence of THz output power on pump fluence, THz PCAs are typically operated at very high fluence near saturation and the damage threshold,\cite{shan2004terahertz} conditions under which responsivities are typically suppressed.

\begin{figure}[!h]
\includegraphics[width=3.37in]{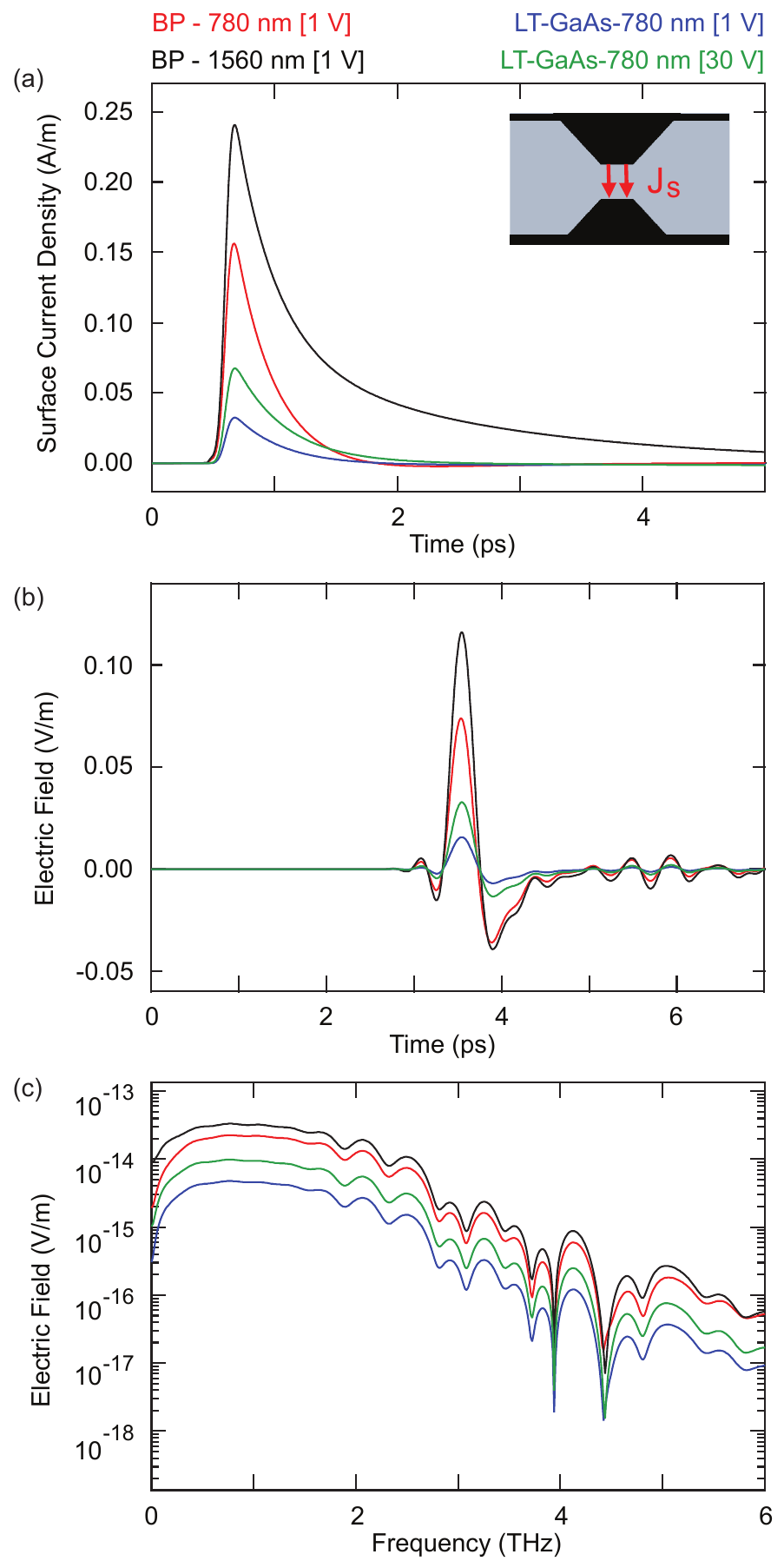}%
\caption{\label{fig5}Results of the computational electromagnetic device model for a BP-based THz PCA with $V_b=1$ V excited at 780 nm (red curves) and 1560 nm (black curves), and for a LT-GaAs-based THz PCA excited at 780 nm with $V_b=1$ V (blue curves) and $V_b=30$ V (green curves). In all cases the average power was 1 mW. (a) Transient surface photocurrent for BP- and LT-GaAs-based PCAs.  (b) THz electric field generated by the surface photocurrent calculated a distance 250 $\mu$m beneath the antenna inside the Si substrate.  (c) Same as (b) in the frequency domain.}%
\end{figure}

We now compare the experimental results with those of the computational model.
Because of the relatively large number of parameters entering the model and the large spread of values reported for BP electrical and optical properties, we seek only qualitative agreement between experiment and the computation model.
The results of Fig.~3(e) and (f) show the photocurrent density calculated as a function of V$_b$. The BP PCA was excited at wavelengths of 780 and 1560 nm, with an average power of 0.43 mW to match the experiment in Fig.~3(b). The results demonstrate similar qualitative behavior to the experimental results of Fig. 3(b), although the amplitudes are observed to be different. This could be due to using the upper limits of the mobility in the model as an effort to estimate the highest potential photocurrent in the BP PCA.
The supplementary Fig.~S8 demonstrates that using the lower limits reported in the literature for BP mobility leads to lower photocurrent density, almost 25\% of that at the BP higher reported mobility.\cite{suess2016mid,prakash2017black,long2016achieving}  Furthermore, saturation of photocurrent density, as seen experimentally, was observed in the model only for higher mobility. 

The results of Fig.~3(f) show the photoconductivity based on the slope of the photoconductive current in Fig.~3(e) over the range $\pm$100 mV. The excitation power was varied from 0 to 1 mW. Direct current at zero power in the model was small, indicating a dark resistance of approximately 1 M$\Omega$, significantly larger than the experimental value of 15 k$\Omega$.  
The computational results demonstrate a qualitatively consistent behavior as the experimental results in Fig. 3(c) for excitation at 780 nm, which is not the case for excitation at 1560 nm. 
For this wavelength, the computational model displays a linear power dependence for low powers.  
In contrast, the experimental results suggest a superlinear behavior at low power. 
Similar increases in responsivity were reported in Refs.~\onlinecite{engel2014black} and \onlinecite{wang2019spectral}.
One possibility is that the superlinear behavior arises from the nonlinear optical response of BP,\cite{lu2015broadband} which is not accounted for in the computational model.
While two-photon absorption could account for such an increase, saturated absorption is typically observed for BP at this wavelength and power range.\cite{lu2015broadband} 
More detailed experimental investigation of the dependence of photoconductivity over a wider range of powers will be required to clarify this point.

Next, we examine experimentally the spatial dependence of photocurrent generated by the BP PCA using scanning photocurrent microscopy, again at excitation wavelengths of 780 and 1560 nm for $V_b=\pm100$ mV.
Comparison of simultaneously acquired reflection and photocurrent images (Fig.~4) indicates that the maximum photocurrent is produced when the laser spot excites the center of the antenna gap.
The maximum photocurrent is shifted to the right side of the gap because of spatial nonuniformity in either the photoconductivity of the BP or the BP/Au contact resistance.
This nonuniformity is specific to this device and was not present in the other two devices measured (cf.~Fig.~S7).
Images for 780 and 1560 nm excitation are largely similar, except for a higher photocurrent for 1560 nm excitation consistent with the data in Fig.~3 and slightly lower resolution for the 1560 nm images, as expected.
Line profiles of the current along the gap [Fig.~4(d),(h)] are fit well with a Gaussian lineshape, with full widths at half maximum of 6.4 (6.8) $\mu$m for photocurrent profiles acquired at 780 (1560) nm.
These widths are approximately equal to the width of the antenna gap plus the size of the excitation beam.
Photocurrent images at $V_b=0$ V are presented in Fig.~S6.

Finally, we use the results of the computational surface photocurrent density of the BP device to compare the anticipated performance of BP THz PCAs with those based on more conventional LT-GaAs.
Calculated surface photocurrent densities for BP excited at 780 (1560) nm, $P_{ave} = 1$ mW, and $V_{b} = 1$ V are shown in Fig.~5(a). The BP results are compared with those of conventional GaAs PCAs at the same power with $V_{b} = 1$ and $30$ V. Here we modeled the GaAs PCA using the same electrode geometry shown in Fig.~1 on top of a LT-GaAs layer of thickness 500 nm. Because the BP and electrodes have the same width in the model, the surface photocurrent densities of all antennas are obtained by multiplying the volume current densities $\textbf{J}(r,t)$ from eq.~5 by the thickness of the BP layer (40 nm). The THz electric field signal $\textbf{E}_{THz}$ is computed in the Si substrate at a distance of 250 $\mu$m beneath the gap for all considered antennas. We measured the complex relative permittivity of Si to be $11.5 - 0.0047i$ using THz time-domain spectroscopy (0.1-3.5 THz), consistent with Ref.~\onlinecite{bolivar2003measurement}. The results of the BP at 780 (1560) nm are compared with those of the conventional GaAs PCAs demonstrating that BP PCA is capable of generating THz signal similar to GaAs as shown in Fig.~5(b). The results also demonstrate that the signal generated at 1560 nm is higher than that at 780 nm, consistent with the larger photocurrent obtained experimentally and computationally. The spectrum of the photocurrents and the THz signals are plotted as a function of frequency in Fig.~5(c), showing a bandwidth of approximately 4 THz for all PCAs with higher amplitude of the BP at 1560 nm. Here we define the bandwidth as the frequency above which the spectrum becomes irregular.  The observed sharp dips in THz electric field spectrum of all PCAs around 4 and 4.5 THz are resonances caused by the dimensions of the antenna. At these frequencies, the dimensions become close to half of a wavelength in the Si substrate.  This issue can be resolved by optimizing the geometry of the antenna metal.

\section{Conclusion}
We have integrated hBN-encapsulated BP into PCAs and characterized their photoresponse at the two most common wavelengths available from low-cost fs fiber lasers, 780 and 1560 nm. 
By engineering the flake thicknesses to create an anti-reflection cavity, absorption by the device reached 36\% and photoconductivity of 1 (2) S/cm was observed for excitation at 780 (1560) nm in the linear regime of bias.
The computational model reproduced the essential qualitative characteristics of the device, including strong absorption by the BP layer, photocurrent saturation, and superior photoconductivity for excitation at 1560 nm compared to 780 nm.
Photocurrent saturation above 1 V and photoconductivity saturation near 1 mW average power show promise for the high-bias, high-power operating regime of THz-emitting PCAs.  Computational results of THz output suggest that BP-based THz PCAs could perform at least as well as those based on GaAs, but we note significant uncertainty in the parameters used as inputs to the model, in particular the mobilities and lifetimes.  

Flake thicknesses in these devices were optimized for absorption at 780 nm; in future devices, significantly better performance at 1560 nm could be achieved by optimizing the stack to absorb at that wavelength (cf.~Fig.~S4).
Additional improvements could be achieved by incorporating a reflective back gate beneath the BP to suppress absorption by Si and control carrier density, by characterizing and optimizing carrier lifetime and/or device transit time, and by suppressing dark current using an ultrathin hBN layer between the BP and Au contacts.
The THz performance of BP PCAs will be characterized in the future using THz time-domain spectroscopy to compare BP-based PCAs with those using GaAs, InGaAs, and other materials.

\begin{acknowledgments}
We acknowledge support from NSF under awards DMR-1610126 (device fabrication) and ECCS-1948225 (characterization), the University of Arkansas Chancellor's Innovation Fund, and the Arkansas Biosciences Institute. K.W. and T.T. acknowledge support from the Elemental Strategy Initiative conducted by the MEXT, Japan, Grant Number JPMXP0112101001, JSPS KAKENHI Grant Numbers JP20H00354 and the CREST(JPMJCR15F3), JST.
\end{acknowledgments}

\section*{Supplementary Material}
The Supplementary Material includes additional figures showing: topography of the device contacts, reflection images of two other devices, absorption calculations for excitation at 1560 nm, DC photocurrent measurements of the device presented in Figs.~3 and 4, photocurrent images at zero bias, photocurrent images for a second device, and photocurrent density calculations using mobilities on the high- and low-end of those reported in the literature for BP.

\section*{Data Availability}
The data that support the findings of this study are available from the corresponding author upon reasonable request.


\providecommand{\noopsort}[1]{}\providecommand{\singleletter}[1]{#1}%

\end{document}